\documentclass[aps,pre,amsmath,amsfonts,amssymb,floatfix,superscriptaddress,nofootinbib]{revtex4}
\usepackage{graphicx}  
\usepackage{dcolumn}   
\usepackage{bm}        
\usepackage{amssymb}   
\usepackage{amsmath}   

\usepackage{color}

\newcommand\be{\begin{equation}}
\newcommand\bea{\begin{eqnarray} \nonumber }
\newcommand\ee{\end{equation}}
\newcommand\eea{\end{eqnarray}}

\begin{document}

\title{A fully consistent, minimal model for non-linear market impact}

\author{J.~Donier} \affiliation{Capital Fund Management, 23-25 Rue de l'Universit\'e, 75007 Paris, France.}\affiliation{Laboratoire de Probabilit\'es et Mod\`eles Al\'eatoires, Universit\'e Pierre et Marie Curie (Paris 6).}
\author{J.~Bonart} \affiliation{Capital Fund Management, 23-25 Rue de l'Universit\'e, 75007 Paris, France.}\affiliation{CFM-Imperial Institute of Quantitative Finance, Department of Mathematics, Imperial College, 
180 Queen's Gate, London SW7 2RH}
\author{I.~Mastromatteo} \affiliation{Centre de Math\'ematiques Appliqu\'ees, CNRS, UMR7641, Ecole Polytechnique, 91128 Palaiseau, France.}
\author{J.-P.~Bouchaud} \affiliation{Capital Fund Management, 23-25 Rue de l'Universit\'e, 75007 Paris, France.}\affiliation{CFM-Imperial Institute of Quantitative Finance, Department of Mathematics, Imperial College, 
180 Queen's Gate, London SW7 2RH}
\date{\today}

\begin{abstract}
We propose a minimal theory of non-linear price impact based on the fact that the (latent) order book is locally linear, as suggested 
by diffusion-reaction models and general arguments. Our framework allows one to compute the average price trajectory in the presence of a meta-order,
that consistently  generalizes previously proposed propagator models. We account for the universally observed square-root impact law, and 
predict non-trivial trajectories when trading is interrupted or reversed. We prove that our framework is free of price manipulation, and that prices
can be made diffusive (albeit with a generic short-term mean-reverting contribution). Our model suggests 
that prices can be decomposed into a transient ``mechanical'' impact component and a permanent ``informational'' component. 
\end{abstract}

\maketitle

\section{Introduction}

The study of market impact (i.e. the way trading influences prices in financial markets) is arguably among the most exciting current themes in 
theoretical finance, with many immediate applications ranging from trading cost modelling to important regulatory issues. What is the meaning of the market price if the very fact 
of buying (or selling) can substantially affect that price? The  questions above would on their own justify a strong research
activity that dates back to the classic Kyle paper in 1985 \cite{Kyle:1985}. But as often in science, it is the empirical discovery of a genuinely surprising result that
explains the recent spree of activity on the subject (see e.g.~\cite{Almgren:2005,Toth:2011,Farmer:2011,Iacopo:2013,Skachkov:2014} and refs.\ therein). In strong
contrast with the predictions of the Kyle model, market impact appears to be neither {\it linear} (in the traded quantity $Q$) nor {\it permanent}, i.e. 
time independent \cite{Bouchaud:2008}. As now firmly established by many independent empirical studies, the average price change induced by the sequential execution of a total volume $Q$ 
(which we call \emph{meta-order}) appears to follow a sub-linear, approximate $\sqrt{Q}$ law  \cite{Barra:1997,GrinoldKahn,Almgren:2005,Moro:2009,Toth:2011,Iacopo:2013,Gomes:2013,Bershova2013,Brokmann:2014}. 
At the end of the meta-order, impact is furthermore observed
to decay (partially or completely) towards the unimpacted price \cite{Gomes:2013,Bershova2013,Brokmann:2014,Donier:2015}.

Quite strikingly, the square-root law appears to be \emph{universal}, as it is to a large degree independent of details such as the type of contract traded (futures, stocks, options, Bitcoin \cite{Donier:2015}...), 
the geographical position of the market venue (US, Europe, Asia), the time period ($1995 \to 2014$), the maturity of the market (e.g. Bitcoin vs. S\&P500), etc. While the impact of {\it single orders} 
is non universal and highly sensitive to market micro-structure, the impact of \emph{meta-orders} appears to be extremely robust against micro-structural changes. 
For example the rise of high-frequency trading (HFT) in the last ten years seems to have had no effect on its validity (compare Refs.~\cite{Barra:1997,Almgren:2005} 
that uses pre-2004 data with \cite{Toth:2011,Iacopo:2013,Gomes:2013} that use post-2007 data). This universality strongly suggests that simple, ``coarse-grained'' models should be
able to reproduce the square-root impact law and other slow market phenomena, while abstracting away from many microscopic details that govern order 
flow and price formation at high frequencies. This line of reasoning is very similar to many situations in physics, where universal large scale/low frequency 
laws appear for systems with very different microscopic behaviour. A well known example is the behaviour of weakly interacting molecules which
on large length scales can be accurately described by the Navier-Stokes equation, with a single ``emergent'' parameter (the viscosity) that encodes the
microscopic specificities of the system. The Navier-Stokes equation can in fact be derived either from the statistical description of the dynamics of molecules, 
through an appropriate coarse-graining procedure, or from general considerations based on symmetries, conservation laws and dimensional 
arguments. Along this path, two pivotal ideas have recently emerged. One is the concept of a {\it latent} order book \cite{Toth:2011} 
that contain the intentions of low-frequency actors at any instant of time, which may or may not materialize in the observable order book. Indeed, since the square-root impact is an aggregate, 
low-frequency phenomenon, the relevant object to consider cannot be the ``revealed'' order book, which chiefly reflects the activity of high frequency market-makers. 
Simple orders of magnitude confirm that the latent liquidity is much higher than the revealed liquidity: whereas the total 
daily volume exchange on a typical stock is around $1/200$th of its market capitalisation, the volume present in the order book at any instant in time is $1000$ times smaller than this. 
Market-makers only act as small intermediaries between much larger volume imbalances present in the latent order book, that can only get resolved on large time scales.  

The second idea is that the dynamics of the latent order book can be faithfully modelled by a so-called ``reaction-diffusion'' model, at least in a
region close to the current price where this dynamics becomes universal, i.e. independent of the detailed setting of the model -- and hence, as 
emphasized above, of the detailed micro-structure of the market and of its high-frequency activity. The reaction-diffusion model in one dimension 
posits that two types of particles (called $B$ and $A$), representing in a financial context the intended orders to buy (\emph{bids}) and to sell 
(\emph{ask}) diffuse on a line and 
disappear whenever they meet $A+B \to \emptyset$ -- corresponding to a transaction. The boundary between the $B$-rich region and the $A$-rich region therefore corresponds to the price $p_t$. 
This highly stylized order book model was proposed in the late 90's by Bak et al.~\cite{Bak:1997,Tang:1999} but never made it to the limelight because the resulting price dynamics 
was found to be strongly mean-reverting on all time scales, at odds with market prices which, after a short transient, behave very much like random walks. 
However, some of us (together with B. T\'oth, \cite{Iacopo:2014}) recently realized that the analogue of market impact can be defined and computed within this framework, 
and was found to obey the square-root law exactly. 

This opens the door to a fully consistent theoretical model of non-linear impact in financial markets, which we propose in the present 
paper. We show how all the previously discussed ingredients can be accommodated in a unifying coarse-grained model for the dynamics of the latent order
book that is consistent with price diffusion, with a single emergent parameter -- the market {\it liquidity} ${\cal L}$, defined below.  When fluctuations are neglected (in a sense that will be specified below), the impact of a meta-order can
be computed exactly, and is found to exhibit two regimes: when the execution rate is sufficiently slow, the model becomes identical to the 
linear propagator framework proposed in \cite{Bouchaud:2004,Bouchaud:2008}, with a bare propagator decaying as the inverse square-root of time. 
When execution is faster, impact becomes fully non-linear and obeys a non-trivial, closed form integral equation. In the two regimes, the impact of
a meta-order grows as the square-root of the volume, but with a pre-factor that depends on the execution rate in the slow regime, but becomes independent
of it in the fast regime -- as indeed suggested by empirical data. The model predicts interesting price trajectories when trading is interrupted or reversed, leading to effects that are observed empirically 
but impossible to account for within a linear propagator model. We demonstrate that prices in our model cannot be 
manipulated, in the sense that any sequence of buy and sell orders that starts and ends with a zero position on markets leads to a non-positive average profit. This is 
a non trivial property of our modelling strategy, which makes it eligible for practical applications. 
Finally, we discuss how our framework suggests a clear separation between ``mechanical'' price moves (i.e. induced by the impact of random trades) and ``informational'' 
price moves (i.e. the impact of any public information that changes the latent supply/demand). This is a key point that allows us to treat consistently, within the same model, 
diffusive prices and memory of the order book -- which otherwise leads to strongly mean-reverting prices (see the discussion in \cite{Toth:2011,Iacopo:2013,Taranto:2014}).
We discuss in the conclusion some of the many interesting problems that our modelling strategy 
leaves open -- perhaps most importantly, how to consistently account for order book fluctuations that are presumably at the heart of liquidity crises and market crashes. 

\section{Dynamics of the latent order book}
\label{dynamics}
Our starting point is the zero-intelligence model of Smith et al.~\cite{Farmer:2003}, reformulated in the context of the latent order book in \cite{Toth:2011} and 
independently in \cite{Lehalle:2010}. 
We assume that each trading (buy/sell) intention of market participant is characterized by a reservation price and a volume.\footnote{
In fact, each participant may have a full time dependent supply/demand curve with different prices and volumes, with little change in the 
effective model derived below.} In the course of time, the dynamics of intentions can be essentially of four
types: a) reassessment of the reservation price, either up or down; b) partial or complete cancellation of the intention of buying/selling; c) appearance of 
new intentions, not previously expressed and finally d) matching of an equal volume of buy/sell intentions, resulting in a transaction at a 
price that delimits the buys/sells regions, and removal of these intentions from the latent order book. It is clear that provided very weak 
assumptions are met: i) the changes in reservation prices are well behaved (i.e. have a finite first and second moment) and short-ranged correlated in time; 
ii) the volumes have a finite first moment, one can establish -- in the large scale, low frequency, ``hydrodynamic'' limit -- the following set 
of partial differential equations for the dynamics of the {\it average} buy (resp. sell) volume density $\rho_B(x,t)$ (resp. $\rho_A(x,t)$) 
at price level $x$:
\begin{eqnarray}
\label{eq:dynamics}\nonumber
\frac{\partial \rho_B(x,t)}{\partial t} &=& - V_t \frac{\partial \rho_B(x,t)}{\partial x} + D \frac{\partial^2 \rho_B(x,t)}{\partial x^2} - 
\nu \rho_B(x,t) + \lambda \Theta(p_t - x)  - \kappa R_{AB}(x,t); \qquad \text{(1-a)} \\ \nonumber
\frac{\partial \rho_A(x,t)}{\partial t} &=& \underbrace{- V_t \frac{\partial \rho_A(x,t)}{\partial x} + 
D \frac{\partial^2 \rho_A(x,t)}{\partial x^2}}_{\text{a- Drift-Diffusion}} - 
\underbrace{\nu \rho_A(x,t)}_{\text{b- Cancel.}} + \underbrace{\lambda \Theta(x - p_t)}_{\text{c- Deposition}} - 
\underbrace{\kappa R_{AB}(x,t)}_{\text{d- Reaction}}; \qquad \,\, \text{(1-b)}
\end{eqnarray}
where the different terms in the right hand sides correspond to the four mechanisms a-d, on which we elaborate below, and 
$p_t$ is the {\it coarse-grained} position of the price (i.e. averaged over high frequency noise), defined from the condition 
\be\label{eq:price-def}
\rho_A(p_t,t) - \rho_B(p_t,t)=0.
\ee 

\begin{itemize}

\item a- {\it Drift-Diffusion:} The first two terms model the fact that each agent reassess his/her reservation price $x$ due to many
external influences (news, order flow and price changes themselves, other technical signals, etc.). One can therefore expect 
that price reassessments contain both a (random) agent specific part that contributes to the diffusion coefficient \footnote{\label{fnt:diff_const}In full generality, 
the diffusion constant $D$ could depend on the distance $|x - p_t|$ to the transaction price. We neglect this possibility in the 
present version of the model, for reasons that will become clear later: see Appendix B. A similar remark applies to the cancellation rate $\nu$ 
as well.} $D$ and a {\it common} component $V_t$ that shifts the entire latent order book. This shift is due to a collective price reassessment due for example 
to some publicly available information (that could well be the past transactions themselves). The drift component $V_t$ is at this stage very general; one possibility that we will 
adopt below is to think of $V_t$ as a white noise, such that the price $p_t$ is a diffusive random walk. Since the derivation of these first two terms
and the assumptions made are somewhat subtle, we devote Appendix A to a more detailed discussion and alternative models; see in particular Eq.~(\ref{eq:Jonathan}).

\item b- {\it  Cancellations:} The third term corresponds to partial or complete cancellation of the latent order, with a decay time $\nu^{-1}$ independent of the
price level $x$ (but see previous footnote$^2$). Consistent with the idea of a common information, cancellation could be correlated between different 
agents. However, this does not affect the evolution of the average densities $\rho_{B,A}(x,t)$, while it might play a crucial role for the 
fluctuations of the order book, in particular to explain liquidity crises. 

\item c- {\it  Deposition:} The fourth term corresponds to the appearance of new buy/sell intentions, modelled by a ``rain intensity'' $\lambda$ 
modulated by an arbitrary increasing function $\Theta(u)$, expressing that buy orders mostly appear below the current price $p_t$ 
and sell orders mostly appear above $p_t$. The detailed shape 
of $\Theta(u)$ actually turns out to be, to a large extent, irrelevant for the purpose of the present paper (see Appendix B for details); for 
simplicity we will choose below a step function, $\Theta(u>0)=1$ and $\Theta(u<0)=0$.

\item d- {\it  Reaction:} The last term corresponds to transactions when two orders meet with ``reaction rate'' $\kappa$; the quantity $R_{AB}(x,t)$ is formally the 
average of the product of the density of $A$ particles and the density of $B$ particles, i.e. $R_{AB}(x,t) \approx \rho_A(x,t)\rho_B(x,t)$ + fluctuations. 
However, the detailed knowledge of $R_{AB}(x,t)$ will not affect the following discussion. We will consider in the following the limit $\kappa \to \infty$, 
which corresponds to the case where latent limit orders close to the transaction price {\it all become instantaneously visible} limit orders that 
are duly executed against incoming market orders.   
\end{itemize}

\begin{figure}[h]
    \begin{center}
        \includegraphics{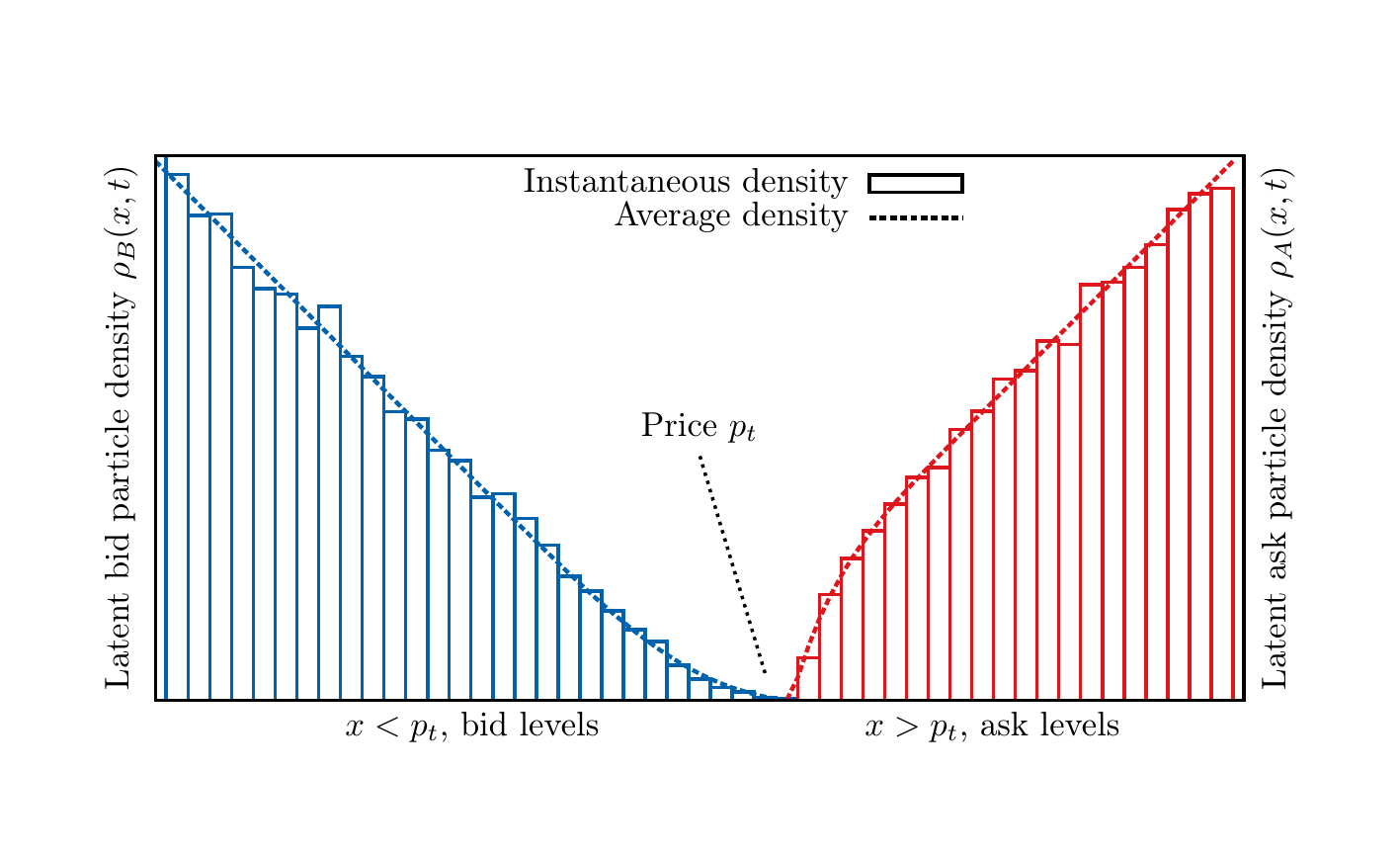}
    \end{center}
    \caption{Snapshot of a latent order book in the presence of a meta-order, with bid orders (blue boxes) and ask orders (red boxes) sitting on opposite sides of the price line and subject to 
    a stochastic evolution. The dashed lines show the mean values the order densities $\rho_{A,B}(x,t)$, which are controlled by Eqs.~~(\ref{eq:dynamics}).}
    \label{fig:ob_scheme}
\end{figure}

Let us insist that Eqs. (\ref{eq:dynamics}-a,b) only describe the {\it average} shape of the latent order book, i.e. fluctuations coming 
from the discrete nature of orders are neglected at this stage: see Fig. \ref{fig:ob_scheme} for an illustration. 
In particular, the instantaneous position of the price $p_t^{\text{inst.}}$ -- where the density
of buy/sell orders vanishes -- has an intrinsic non-zero width even in the limit $\kappa \to \infty$ \cite{Cardy:1996}, corresponding to the average distance $a-b$ between the highest 
buy order $x=b$ and the lowest sell order $x=a$.\footnote{We indeed assume that latent orders become instantaneously visible when close to $p_t^{\text{inst.}}$, 
in such a way that the latent order book and the observable order book become identical at the best limits. This is of course needed to identify $p_t^{\text{inst.}}$ with the 
``real'' mid-price. It is very interesting to ask what happens if the conversion speed between latent orders and real orders is not infinitely fast, or when market orders become
out-sized compared to the prevailing liquidity. As we discuss in the conclusion, this is a potential mechanism for crashes, and the simple coarse-grained framework discussed here has to be
adapted to deal with these situations.}  The instantaneous price can then be conventionally be defined as 
$p_t^{\text{inst.}}=(a+b)/2$, but will in general not coincide with the coarse-grained price 
$p_t$ defined by the average shape of the latent order book through Eq.~(\ref{eq:price-def}). Indeed, as shown in \cite{Cardy:1996}, 
the diffusion width (i.e. the typical distance between $p_t^{\text{inst.}}$ and 
$p_t$) is also non zero and actually larger than the intrinsic width, but only by a logarithmic factor.

In the following, we will neglect both the intrinsic width and the diffusion width, which is justified if we focus on price changes much larger than these widths. 
This is the large scale, low frequency regime where our coarse-grained equations Eqs. (\ref{eq:dynamics}-a,b) are warranted. Formally, Eqs. (\ref{eq:dynamics}-a,b) become valid when the market 
{\it latent liquidity} $\cal L$ (defined below) tends to infinity, since both the intrinsic width and the diffusion width vanish as ${\cal L}^{-1/2}$ \cite{Cardy:1996}.

\section{Stationary shape of the latent order book}

A remarkable feature of Eqs. (\ref{eq:dynamics}-a,b) is that although the dynamics of $\rho_A$ and $\rho_B$ is non-trivial because of the reaction term 
(that requires a control of fluctuations, see \cite{Cardy:1996}) the combination $\varphi(x,t):= \rho_B(x,t) - \rho_A(x,t)$ evolves according to a linear equation independent of 
$\kappa$:\footnote{The disappearance of $\kappa$ can be traced to the conservation of $\# A - \#B$ for each reaction $A+B \to \emptyset$.} 
\begin{equation}
\label{eq:diff_psi}
\frac{\partial \varphi(x,t)}{\partial t} = - V_t \frac{\partial \varphi(x,t)}{\partial x} + 
D \frac{\partial^2 \varphi(x,t)}{\partial x^2} - \nu \varphi(x,t) + \lambda \, \text{sign}\left(p_t - x\right),
\end{equation}
where $p_t$ is the solution of $\varphi(p_t,t)=0$. This solution is expected to be unique for all $t > 0$ if it is unique at $t=0$ (see also \cite{Lehalle:2010}).
Note that Eq.~(\ref{eq:diff_psi}), without the drift-diffusion terms, has recently been obtained as the hydrodynamic limit of a Poisson order book dynamics in \cite{hydro:2014}.

Introducing $\widehat p_t = \int_0^t {\rm d}s \, V_{s}$, the above equation can be rewritten in the reference 
frame of the latent order book $y= x - \widehat p_t$ as:\footnote{One should be careful with the ``Ito'' term when $V_s$ is a Wiener noise, which adds
a contribution to $D$, see Appendix A and Eq.~(\ref{eq:J2})}
\begin{equation}
\label{eq:diff_psi2}
\frac{\partial  \varphi(y,t)}{\partial t} =  D \frac{\partial^2 \varphi(y,t)}{\partial y^2} - \nu \varphi(y,t) + 
\lambda \, \text{sign}\left(p_t  - \widehat p_t - y\right).
\end{equation}
Starting from a symmetric initial condition $\varphi(y,t=0)=-\varphi(-y,t=0)$ such that $p_{t=0}=\widehat p_{t=0}=0$, it is clear by symmetry that the 
equality $p_t = \widehat p_t$ is a solution at all times, since all terms in the above equation are odd when $y \to -y$. For more general initial conditions, 
$p_t$ converges to $\widehat p_t$ when $t \to \infty$ and the stationary solution of Eq.~(\ref{eq:diff_psi2}) reads, in the 
limit $\mu \to \infty$:
\begin{equation}
\label{eq:stat}
\varphi_{\text{st.}}(y \leq 0) = \frac{\lambda}{\nu} \left[1 - e^{\gamma y}\right]; \qquad \varphi_{\text{st.}}(y \geq 0) = -  \varphi_{\text{st.}}(-y),
\end{equation}
with $\gamma^2 = \nu/D$. This is precisely the solution obtained in \cite{Toth:2011} which behaves linearly close to the transaction price.  But as
emphasized in \cite{Toth:2011} and in Appendix B, this linear behaviour in fact {\it holds for a very wide range of models} -- for example if the appearance of new orders only takes 
place at some arbitrary boundary $y = \pm L$, as in \cite{Iacopo:2014}, or else if the coefficients $D, \nu$ are non-trivial (but sufficiently regular) functions of the distance to the price $|y|$, etc.

\section{Price dynamics within a locally linear order book (LLOB)} 

We will therefore, in the following, ``zoom'' into the universal linear region by taking the formal limit $\gamma \to 0$ with a fixed current 
\be
J = D |\partial_y \varphi_{\text{st.}}|_{y=0} \equiv \lambda/\gamma.
\ee
This current can be interpreted as the volume transacted per unit time in the stationary
regime, i.e. the total quantity of buy (or sell) orders that get executed per unit time. As a side remark, it is important to realize that if the drift $V_t$ contains a Wiener noise component, or jumps, this
drift does in fact contribute to $J$ and does not merely shift the latent order book around without any transactions (see Appendix A). 

In the limit $\nu, \lambda \to 0$ with $\lambda/\gamma = J$ fixed, the stationary solution 
$\varphi_{\text{st.}}(y)$ 
becomes exactly linear: 
\be
\varphi_{\text{st.}}(y) = -Jy/D.
\ee 
This is the regime we will explore in the present paper, although we will comment below on the expected modifications induced by non-zero values of $\nu, \lambda$. Note that 
${\cal L} = J/D \equiv \lambda \sqrt{D/\nu}$ can be interpreted as the {\it latent liquidity} of the market, which is large when deposition of latent orders is intense ($\lambda$ large) and/or when 
latent orders have a long lifetime ($\nu$ small). The quantity ${\cal L}^{-1}$ is the analogue, within a LLOB, of Kyle's ``lambda'' for a flat order book.  

In terms of order of magnitudes, it is reasonable to expect that the latent order book has a memory time $\nu^{-1}$ of several hours to several days \cite{Toth:2011} 
-- remember that we are speaking here of slow actors, not of market makers contributing to the high-frequency dynamics of the 
revealed order book. Taking $D$ to be of the order of the price volatility, the width of the linear region $\gamma^{-1}$ is found to be of the order of $1 \%$ of the price 
(see Eq.~(\ref{eq:stat})).  Therefore, we expect that restricting the analysis to the {\it linear} region of the order book will be justified for meta-orders lasting up to several hours, and impacting the 
price by less than a fraction of a percent. For larger impacts and/or longer execution times, a more elaborate (and probably less universal) description may be needed.

We now introduce a ``meta-order'' within our framework and work out in detail its impact on the price. Working in the reference frame of the {\it unimpacted} price $\widehat p_t$ defined 
above, we model a meta-order as an extra current of buy (or sell) orders that fall exactly on the transaction price $p_t$. Introducing $y_t \equiv p_t - \widehat p_t$, the corresponding 
equation for the latent order book reads, within a LLOB that precisely holds when $\nu, \lambda \to 0$:
\begin{equation}
\label{eq:diff_psi3}
\left\{
\begin{aligned}
&\frac{\partial  \varphi(y,t)}{\partial t} =  D \frac{\partial^2 \varphi(y,t)}{\partial y^2} + m_t \delta(y-y_t)\\
&\frac{\partial  \varphi(y \to \pm \infty, t)}{\partial y} = -{\cal L},
\end{aligned}
\right.
\end{equation}
where $m_t$ is the (signed) trading intensity at time $t$; $m_t > 0$ corresponding to a buy meta-order. Note that the
meta-order will be assumed to be small enough not to change the behaviour of the rest of the market (i.e. the parameters $D$, $\nu$ and $\lambda$), 
so that ${\cal L}$ is a fixed parameter in the above equation. Of course, this assumption might break down when the meta-order is out-sized, leading to a sudden increase of the cancellation rate $\nu$ and a corresponding drop of the liquidity ${\cal L}$, which might in turn result in a crash (see the 
discussion in the conclusion).

We will now consider a meta-order that starts at a random time that we choose as $t=0$, with no information on the state of the latent order book. This means that at $t=0$, there is no conditioning on the state of the order book 
that can be described by its stationary shape,
$\varphi_{\text{st.}}(y) = -Jy/D$. For $t > 0$, the
latent order book is then given by the following exact formula:
\begin{equation}
\label{eq:sol_diff_psi3}
\varphi(y,t) =   - {\cal L} y  + \int_0^t \frac{{\rm d}s\, m_s}{\sqrt{4 \pi D (t-s)}} e^{-\frac{(y - y_s)^2}{4D(t-s)}},
\end{equation}
where $y_s$ is the transaction price (in the reference frame of the book) at time $s$, defined as $\varphi(y_s,s) \equiv 0$. This leads to a self-consistent integral equation for the price at time $t > 0$:
\begin{equation}
\label{eq:price}
y_t  = \frac{1}{{\cal L}} \int_0^t \frac{{\rm d}s\, m_s}{\sqrt{4 \pi D (t-s)}} e^{-\frac{(y_t - y_s)^2}{4D(t-s)}}.
\end{equation}
This is the central equation of the present paper, which we investigate in more detail in the next sections.\footnote{When $m_s$ has a non-trivial time dependence, the above equation may not 
be easy to deal with numerically. It can be more convenient to iterate numerically the Eq.~(\ref{eq:diff_psi3}) and find the solution of $\varphi(y_t,t)=0$.}

As a first general remark, let us note that provided impact is small, in the sense that $\forall t,s$, $|y_s - y_t|^2 \ll D (t-s)$,  then the above formula exactly boils down 
to the {\it linear propagator model} proposed in \cite{Bouchaud:2004,Bouchaud:2008} (see also \cite{Gatheral:2010}), with a square-root decay of impact:
\begin{equation}
\label{eq:prop}
y_t  = \frac{1}{{\cal L}} \int_0^t \frac{{\rm d}s\, m_s}{\sqrt{4 \pi D (t-s)}}.
\end{equation}
This linear approximation is therefore valid for very small trading rates $m_s$, but breaks down for more aggressive executions, for which a more precise analysis is needed. An ad-hoc non-linear generalisation
of the propagator model was suggested by Gatheral \cite{Gatheral:2010}, but is difficult to justify theoretically (and 
leads to highly singular optimal trading schedules in the continuous time limit \cite{Lillo:2015}). 
We believe that Eq.~(\ref{eq:price}) above is 
the correct way to generalize the propagator model, such that all known empirical results can be qualitatively accounted for.

Note that one can in fact define a volume dependent ``bid'' (or ``ask'') price $y_t^{\pm}(q)$ for a given volume $q$ as the solution of:
\be
\int_{y_t^-(q)}^{y_t} {\rm d}y \,  \varphi(y,t) = - \int_{y_t}^{y_t^+(q)} {\rm d}y \,  \varphi(y,t) = q.
\ee
Clearly, in the equilibrium state, and for $q$ small enough, $y_t^{\pm}(q) = y_t \pm \sqrt{2q/{\cal L}}$. After a buy meta-order, however, we will find that strong asymmetries can appear.

\begin{figure}[h]
    \begin{center}
        \includegraphics{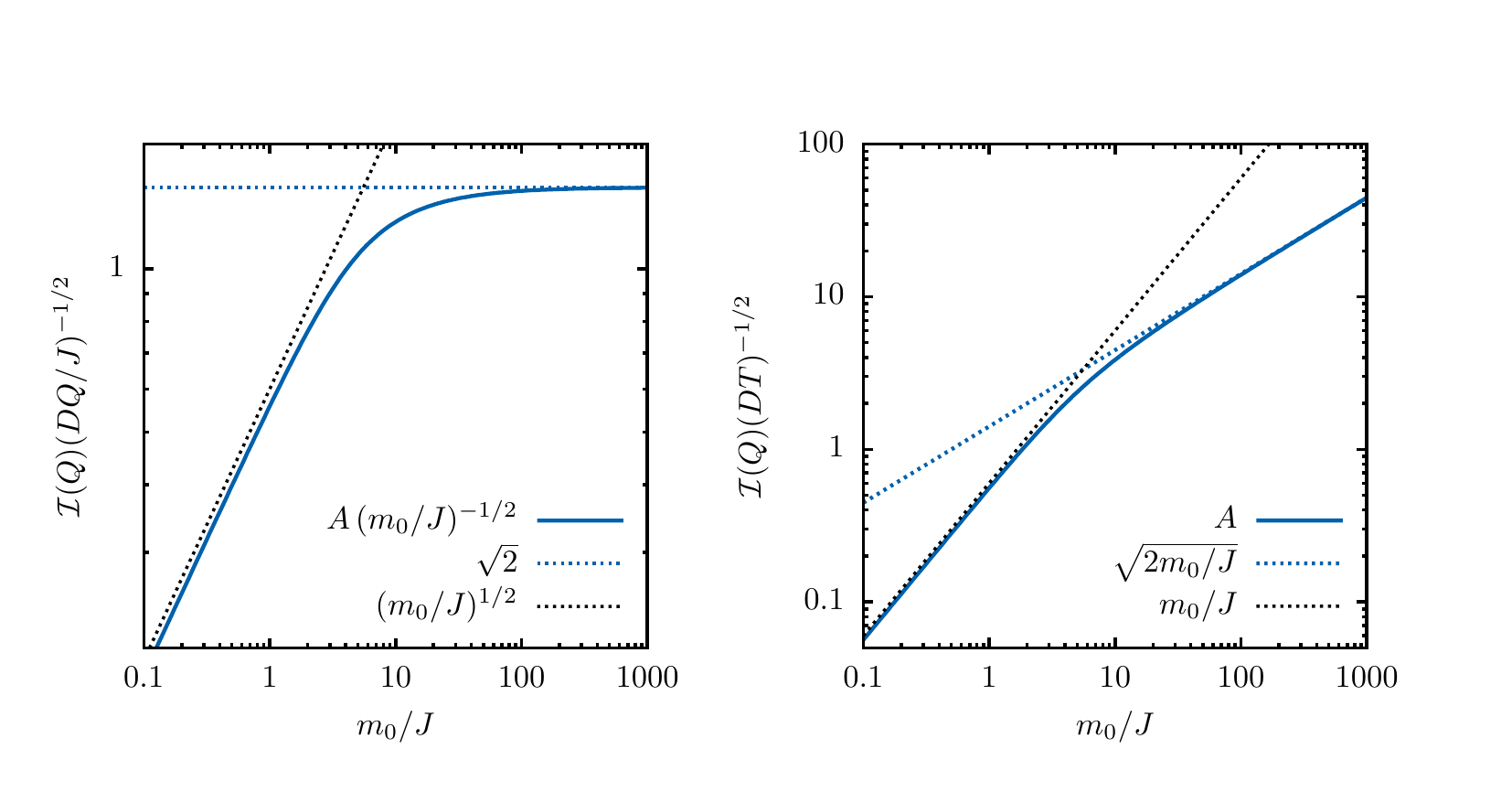}
    \end{center}
    \caption{Left: Dependence of the ratio $A/\sqrt{m_0/J}$ upon the trading rate parameter $m_0/J$. (This ratio coincides with the empirically used $Y$ ratio if 
    $\sigma^2$ is identified with $D$ and $V$ with $J$).  
    The curve interpolates between a $\sqrt{m_0/J}$ dependence observed at small trading trading rates and an asymptotically constant 
    regime $\approx \sqrt{2}$ for large $m_0/J$. This is consistent with the weak dependence of $Y$ upon the trading rate observed in CFM empirical data. Right: Dependence of the 
    impact ${\cal I}(Q)$ on $Q$ for a {\it fixed} execution time $T$ -- i.e. a variable $m_0=Q/T$. Note the crossover between a linear behaviour at small $Q$ and a square-root behaviour for 
    large $Q$.}
    \label{fig:m_vs_y_ratio}
\end{figure}

\section{The square-root impact of meta-orders}

The simplest case where a fully non-linear analysis is possible is that of a meta-order of size $Q$ executed at a constant rate $m_0=Q/T$ for $t \in [0,T]$. In this case, 
it is straightforward to check that $y_s = A \sqrt{Ds}$ is an {\it exact} solution of Eq.~(\ref{eq:price}), where the constant $A$ is the solution of the following equation:
\begin{equation}
A = \frac{m_0}{J} \int_0^1 \frac{{\rm d}u}{\sqrt{4 \pi  (1-u)}} e^{-\frac{A^2(1 - \sqrt{u})}{4(1 + \sqrt{u})}}.
\end{equation}
It is easy to work out the asymptotic behaviour of $A$ in the two limits $m_0 \ll J$ and $m_0 \gg J$. In the first case, one finds $A \approx m_0/J \sqrt{\pi}$, while in the second case $A \approx \sqrt{2m_0/J}$. 
The impact ${\cal I}$ of a meta-order of size $Q$, is defined as:\footnote{Note that ${\cal I}(Q)$ is a slight abuse of notations since the impact
in fact depends in general in the whole trajectory $m_s$.}
\be
{\cal I}(Q) = \langle \varepsilon \cdot (p_{t+T} - p_t) |Q \rangle,
\ee
where $\langle \dots |Q \rangle$ denotes an average over all meta-orders of sign $\varepsilon$ and volume $Q$, executed over the time interval $[t,t+T]$.

We assuming for now that the meta-order is uninformed, in the following sense:
\be 
\langle \varepsilon \cdot (\hat p_{t+T} - \hat p_t) |Q \rangle = 0,
\ee
such that the only contribution is the ``mechanical'' impact on the dynamics of $y_t$. The case of informed meta-orders will be treated in Sect. IX.
The mechanical impact at the end of the meta-order is then given by $y_T = A \sqrt{DT}$, i.e.:\footnote{The results in the two limits are (up to prefactors) those obtained in \cite{Iacopo:2014} within an explicit reaction-diffusion setting.}
\be\label{eq:impact}
{\cal I}(Q) = \frac{A}{\sqrt{m_0}} \sqrt{DQ} \approx \sqrt{\frac{m_0}{J \pi}} \times \sqrt{\frac{Q}{{\cal L}}} \qquad (m_0 \ll J); \qquad \qquad{\cal I}(Q)
\approx \sqrt{2\frac{Q}{{\cal L}}} \qquad (m_0 \gg J),
\ee
i.e. precisely a square-root impact law. 

In fact, the empirical result is often written as ${\cal I}(Q) = Y \sigma \sqrt{Q/V}$ 
where $\sigma$ is the daily volatility and $V \equiv JT_d = D{\cal L}T_d$ the daily traded volume ($T_d \equiv 1$ day), and $Y$ a constant of order unity. Assuming that $\sigma^2 \propto D T_d$ (which is the case if $D_0=0$, see Appendix A), we see that Eq.~(\ref{eq:impact}) exactly reproduces the empirical result, with $Y$ proportional to $\sqrt{m_0/J}$ 
for small trading intensity $m_0$ and becoming independent of $m_0$ for larger trading intensity -- see Fig.~\ref{fig:m_vs_y_ratio}.\footnote{Note that in agreement with our interpretation of the latent order book, the quantity 
$JT$ must be interpreted as the volume of ``slow'' orders executed in a time $T$, removing all fast intra-day activity that averages out and therefore cannot withstand (other than temporarily) the incoming meta-order.} CFM's empirical data indeed suggests that $Y$ only very weakly depends on the trading intensity, which is nicely explained by the present framework. 

\section{Impact decay: beyond the propagator model}

The next interesting question is impact relaxation: how does the price behave after the meta-order has been executed, i.e. when $t > T$. Mathematically, the impact decay is given by the solution of:
\begin{equation}
\label{eq:relax}
y_t  = \frac{D m_0}{J} \int_0^T \frac{{\rm d}s}{\sqrt{4 \pi D (t-s)}} e^{-\frac{(y_t - A \sqrt{Ds})^2}{4D(t-s)}},\qquad (t>T)
\end{equation}
In the small $m_0/J$ limit, the linear propagation model is appropriate and predicts the following impact relaxation:
\be\label{eq:decay}
\frac{{\cal I}(Q, t > T)}{{\cal I}(Q)} = \frac{\sqrt{t} - \sqrt{t-T}}{\sqrt{T}},
\ee
that behaves as $1 - \sqrt{(t-T)/T}$ very shortly after the end of the meta-order and as $\sqrt{T/t}/2$ at long times. 

\begin{figure}[h]
    \begin{center}
        \includegraphics{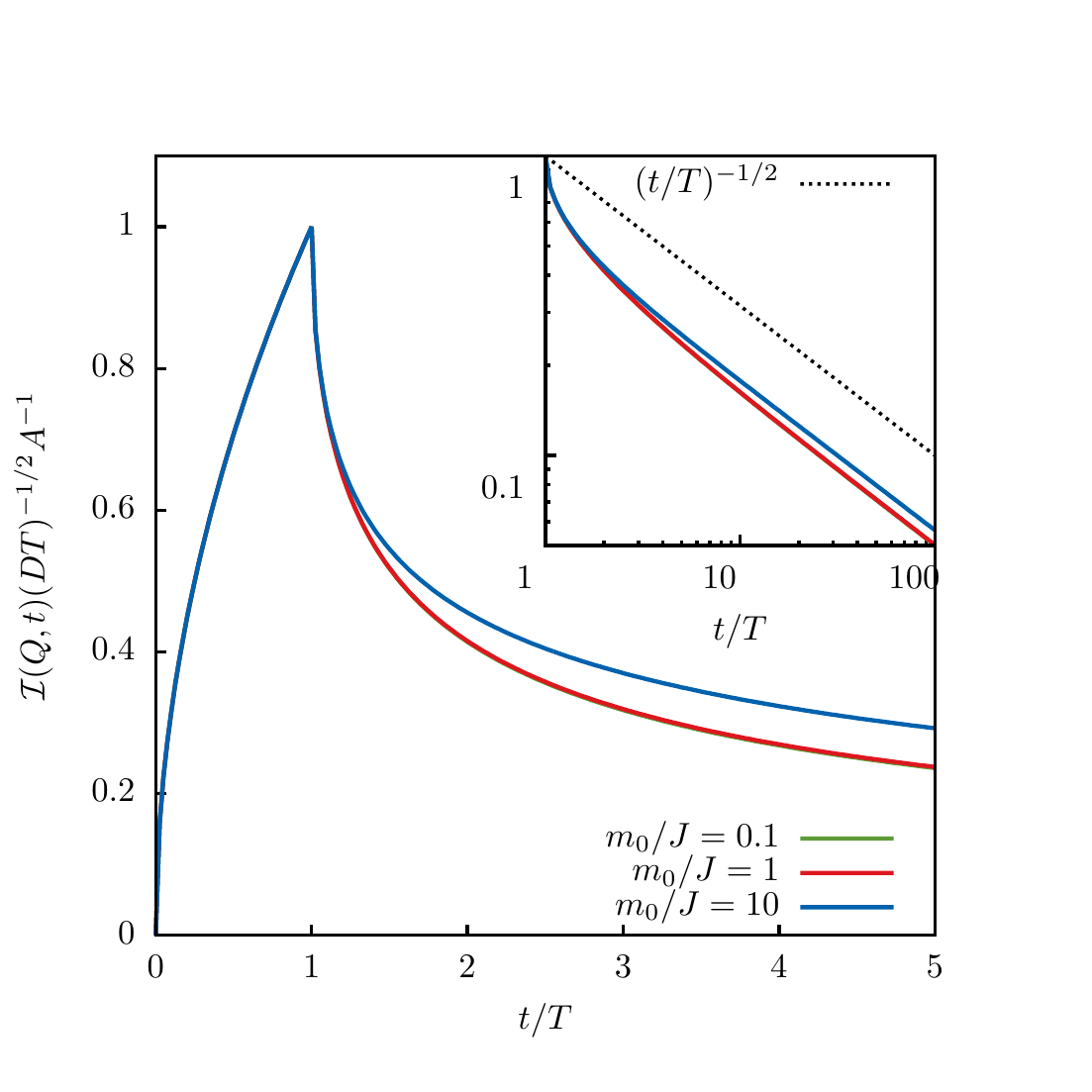}
    \end{center}
    \caption{Impact $\mathcal{I}(Q,t)$ as a function of rescaled time for various trading rate parameters $m_0/J$. 
    The initial growth of the impact follows exactly a square-root law, and is followed by a regime shift suddenly after end of the meta-order.
    While for $t=T^+$ the slope of the impact function becomes infinite, at large times one observes an inverse square relaxation
    $\sim \sqrt{T/t}$ with an $m_0/J$ dependent pre-factor. Note that the curves for $m_0/J=0.1$ and $m_0/J=1$ are nearly indistinguishable.}
    \label{fig:imp_decay}
\end{figure}

\begin{figure}[h]
    \begin{center}
        \includegraphics{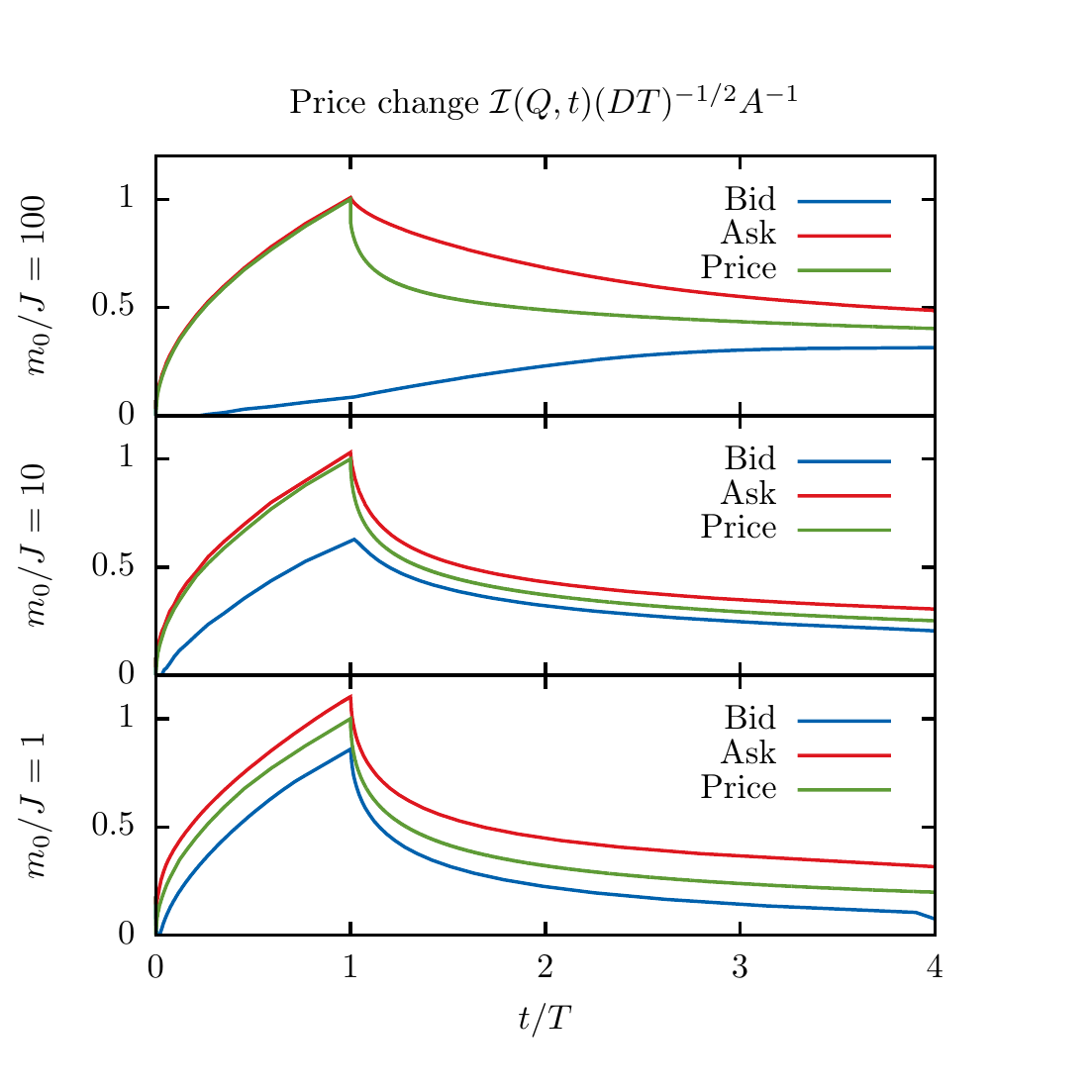}
    \end{center}
    \caption{Evolution in time of the bid $p^-_t(q)$ (blue line) and the ask $p^+_t(q)$ (red line) while executing a meta-order at a rate $m_0/J \in \{1,10,100\}$.
    The price $p_t$ (green line) is also shown for comparison. The three curves correspond to the execution of a constant volume $Q=m_0 T$, 
    while the threshold $q$ has been set by $q=10^{-3} \, Q$. 
    The plot illustrates how a large execution rate $m_0/J$ induces a locally asymmetric liquidity profile around the price, see also Fig.~\protect\ref{fig:full_book_evol}.}
    \label{fig:bid_ask}
\end{figure}

The analysis of Eq.~(\ref{eq:relax}) at large $m_0/J$ is more subtle, in particular at short times. The full analysis is given 
in Appendix C and reveals that the rescaled initial decay of impact is, quite unexpectedly, still 
exactly given by Eq.~(\ref{eq:decay}), independently of $m_0/J$. For large times, $y_t \to 0$, which implies that asymptotically 
$|y_t - A \sqrt{Ds}| \ll \sqrt{Dt}$, i.e. the exponential term in Eq.~(\ref{eq:relax})
is approximately equal to one, leading to an asymptotic rescaled impact decay as $\sqrt{m_0 T/2 \pi Jt}/4$. We plot in 
Fig.~\ref{fig:imp_decay} the normalized free decay of impact for different values of $m_0/J$ for the ``mid-price'' $p_t$, and in Fig. \ref{fig:bid_ask} 
the corresponding evolution of the effective ``bid-ask'' $p^\pm_t(q)$ for a given volume $q$, illustrating how the latent order book becomes more and more
asymmetric as $m_0/J$ increases.

\begin{figure}[h]
    \begin{center}
        \includegraphics{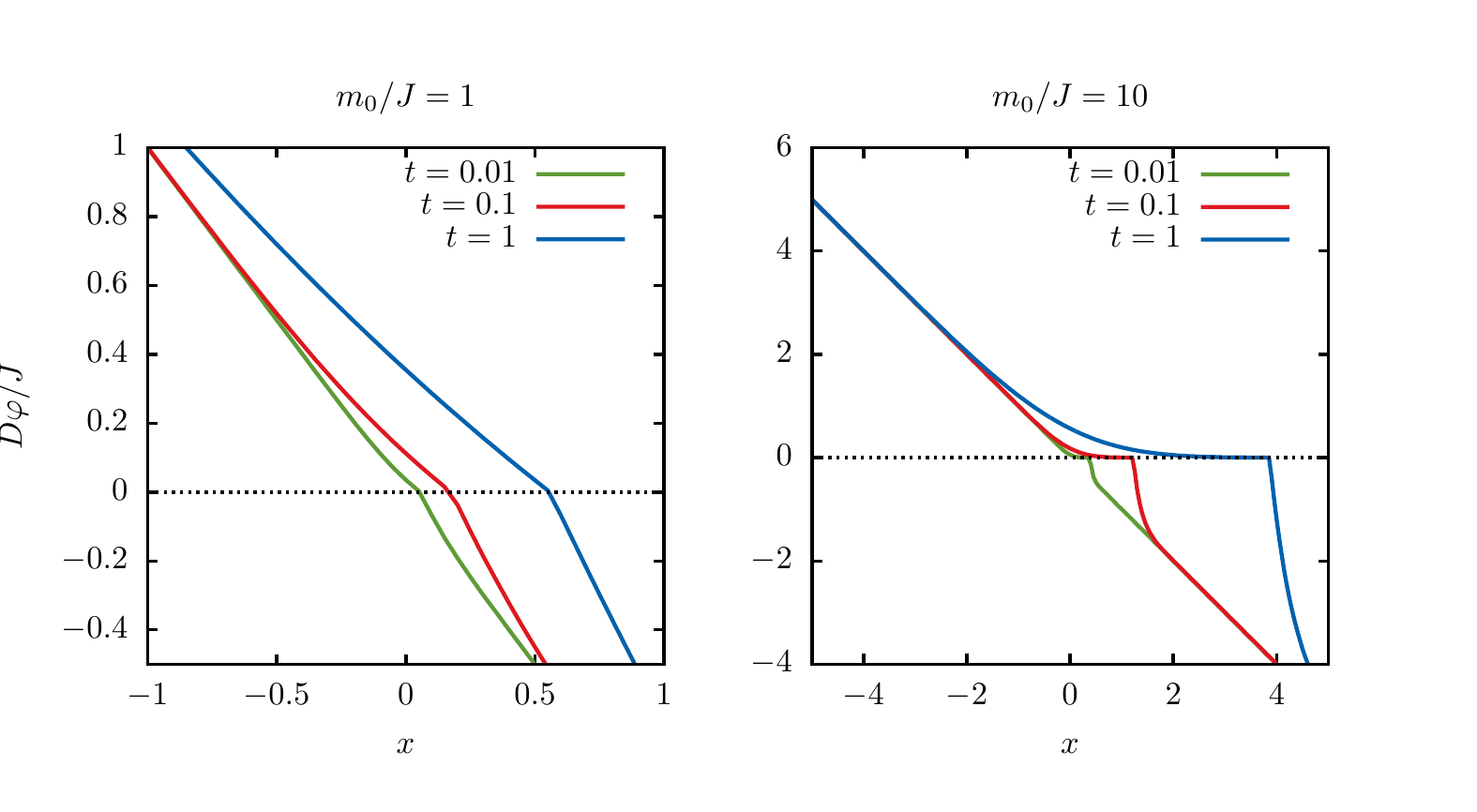}
    \end{center}
    \caption{Evolution of the order book shape $\varphi(x,t)$ during the execution of a meta-order at small trading rate $m_0/J=1$ 
    (left plot) and large trading rate $m_0/J=10$ (right plot). 
    The solid lines indicate the profile of the book at $t=0.01$ (green line), $t=0.1$ (red line) and $t=1$ (blue line). 
    While the displacement of the mid-price follows a square root law, the function $D \varphi(x,t)/J+x$ satisfies 
    a scaling relation determined by the parameter $m_0/J$ -- see also Appendix C and Fig.~\ref{fig:book_evol}.}
    \label{fig:full_book_evol}
\end{figure}

\begin{figure}[h]
    \begin{center}
        \includegraphics{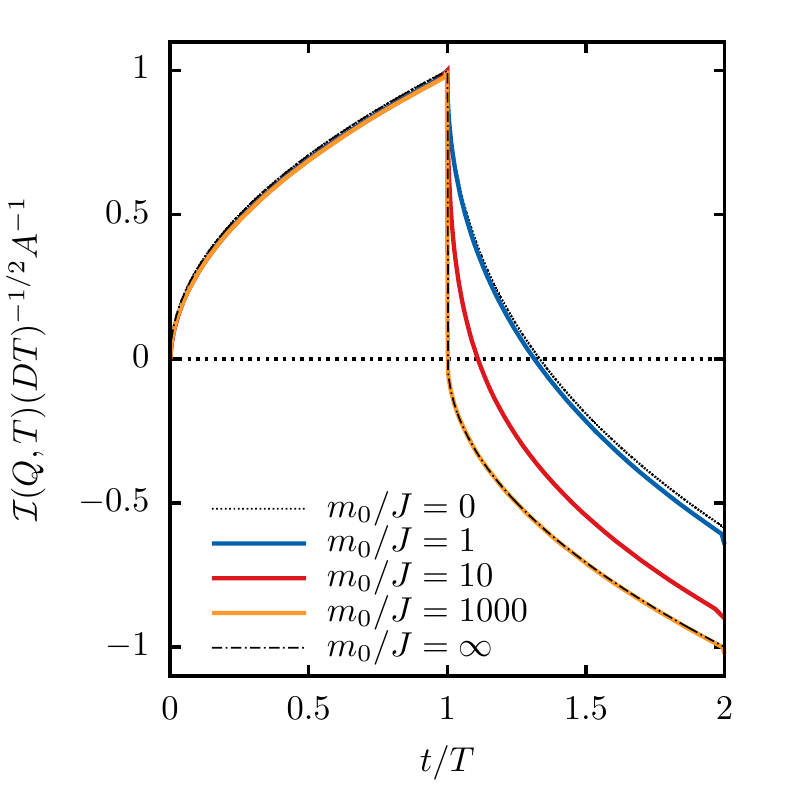}
    \end{center}
    \caption{Trajectory of the average price before and after a sudden switch of the sign of a meta-order. 
    We have considered $m_t=m_0$ for $t<T$ and $m_t=-m_0$ for $t>T$, and plotted the expected price change as a 
    function of time for different values of $m_0$. The curves for finite $m_0$ (solid lines) are 
    also compared with the theoretical benchmark $m_0=0$, corresponding to the propagator model (dotted line), and to the 
    $m_0=\infty$ limit (dot-dashed line). We find that non-linear effects in the large $m_0$ regime makes to propagator approximation 
    invalid, and increase considerably the impact of the reversal trade.}
    \label{fig:price_reversal}
\end{figure}

The above analysis can be extended to the case where trading is reverted after time $T$, i.e. $m_t = m_0$ for $t \in [0,T]$ and $m_t = -m_0$ for $t \in [T,2T]$. 
This case is particularly interesting since it puts the emphasis on the lack of liquidity behind the price for large execution rates. 
Within the linear propagator approximation, it is easy to show that the time needed for the price to come back to its initial value (before continuing to be pushed down by the sell meta-order) is given by $T/4$. 
In the non linear regime $m_0 \gg J$, the price goes down much faster, and reaches its initial value after a time given by $JT/2m_0 \ll T/4$ -- see Figs.~\ref{fig:full_book_evol},~\ref{fig:price_reversal}. 
Such an asymmetry is indeed seen empirically \cite{Benichou}, and means that such a simple round-trip is necessary costly, since the average sell price is below the average buy price. We shall see below that 
this property (called absence of price manipulation in \cite{HubermanStanzl:2001,Schied:2010}) holds in full generality within our framework.

\section{Price trajectory at large trading intensities}

Our general price equation Eq.~(\ref{eq:price}) is amenable to an exact treatment in the large trading intensity limit $m_t \gg J$, provided $m_t$ does not change sign and is a sufficiently regular function of time. 
In such a case, the change of price is large and therefore justifies a saddle-point estimate of the integral appearing in Eq.~(\ref{eq:price}). This leads to the following asymptotic equation of motion:
\be \label{eq:expansion}
{\cal L} y_t |\dot y_t | \approx m_t \left[ 1 + D \left(3 \frac{\ddot y_t}{\dot y_t^3} - 2 \frac{\dot m_t}{m_t \dot y_t^2} \right) + 
O\left(\frac{J^2}{m^2}\right) \right];
\ee 
see Appendix D for details of the derivation and for the next order term, of order $J^2/m^2$.

When $m_t$ keeps a constant sign (say positive), the leading term of the above expansion therefore yields the following average impact trajectory:
\be
y_t  \approx \sqrt{\frac{2}{{\cal L}} \int_0^t {\rm d}s \, m_s},
\ee
i.e. a price impact that only depends on the total traded volume, but not on the execution schedule. This is a stronger result than the one obtained above, where impact was found to be independent 
of the trading intensity for a uniform execution scheme. This path independence is in qualitative agreement with empirical results obtained at CFM. Using Eq.~(\ref{eq:expansion}), systematic corrections to 
the above trajectory can be computed (see Appendix D). Perhaps surprisingly, the execution cost of a given quantity $Q$ is found to be {\it independent} of the trading schedule even to first-order in $J/m$ --
see Appendix D for a proof. Exploring the optimal execution schedule within the full non-linear price equation Eq.~(\ref{eq:price}), and comparing the results with those obtained 
in Ref.~\cite{Lillo:2015}, is left for a future study. 

\section{Absence of price manipulation}

We now turn to a very important issue, that of price manipulation. Although not proven to be impossible in reality, it looks highly implausible that one will ever be able to build a money machine that 
``mechanically'' pumps money out of markets.  Any viable model of price impact should therefore be such that mechanical price manipulation, leading to a positive profit after a closed trading loop, 
is impossible in the absence of information about future prices \cite{HubermanStanzl:2001}.\footnote{Note that property is highly important for practical purposes as well, since using an impact model with 
profitable closed trading trajectories in -- say -- dynamical portfolio algorithms would lead to instabilities.} Here, we show that the non-linear price impact model defined by Eq.~(\ref{eq:price}) is free 
of price manipulation, generalizing the result of \cite{Schied:2010} for the linear propagator model, see also \cite{Blanc:2014}. We start by noticing that the average cost of a closed trajectory is given by:
\be
{\cal C} = \int_0^T {\rm d}s \, m_s y_s, \qquad {\mbox{with}} \quad  \int_0^T {\rm d}s \, m_s  = 0
\ee 
and $y_s$ given by Eq.~(\ref{eq:price}). The above formula simply means that the executed quantity $m_s {\rm d}s$ between time $s$ and $s + {\rm d}s$ is at price $y_s$.\footnote{The alert reader might wonder whether $m_s$ is really the {\it executed} quantity, rather than the {\it submitted} quantity, 
as the definition of $m_s$ as a flux of buy/sell orders suggest. However, within the present framework where $m_s$ is deposited precisely at the 
mid-price $p_t$, one can check that in the limit $\kappa \to \infty$, and provided latent and real liquidity are the same close to $p_t$, the opposite flow of limit orders immediately adapts to absorb exactly the incoming meta-order.} 
Because the initial and final positions are 
assumed to be zero, there is no additional marked-to-market boundary term. Using Eq.~(\ref{eq:price}), it is not difficult to show that ${\cal C}$ can be identically rewritten as a quadratic form:
\be
{\cal C} = \frac12 \int_0^T \int_0^T {\rm d}s {\rm d}s' \, m_s \, M(s,s') \, m_{s'},
\ee
where $M(s,s')$ is a non-negative operator, since it can be written as a sum of ``squares'' $KK^\dagger$, or more precisely:
\be
M(s,s') = \frac{D}{{\cal L}} \int_{-\infty}^{\infty} {\rm d}z \, z^2 \int_{-\infty}^{+\infty} {\rm d}u \, K_z(s,u)K_z^*(s',u), \qquad K_z(s,u) \equiv \Theta(s-u)e^{-Dz^2(s-u) + izy_s}.
\ee
This therefore proves that ${\cal C} \geq 0$ for {\it any} execution schedule, i.e. price manipulation is impossible within a LLOB (see \cite{Skachkov:2014} for loosely related ideas). 
We note, {\it en passant}, that this proof extends to a much larger class of Markovian order book dynamics, where the reservation price of latent orders evolves, for example, according to a L\'evy process (and not necessarily a diffusion, as assumed heretofore -- see Appendix A). 

\section{Mechanical vs. informational impact}

\label{decomp}
We now imagine that the agent executing his/her meta-order has some information about the future price, i.e. that the execution flow $m_t$ is correlated with the future motion of the latent order book 
$V_{t'}$ for $t' > t$. The apparent impact of the meta-order will now contain two contributions that are, within our framework, {\it additive}. 
Assuming again, for simplicity, that $m_t=m_0$, one finds that the average price difference can be written as:
\be
\langle \varepsilon \cdot (p_t - p_0) | Q\rangle = \langle \varepsilon \cdot (\widehat p_t - \widehat p_0) | Q \rangle  + \langle \varepsilon \cdot (y_t - y_0) | Q \rangle,
\ee
where now the first term is non-zero. More explicitly, this leads to:
\be\label{2terms}
\langle p_t - p_0 \rangle = m_0 \int_0^t {\rm d}s \int_0^s {\rm d}s' C(s - s') + A(m_0) \sqrt{Dt}, \qquad (t \leq T)
\ee
where $C(s-s') \propto \langle V_{s} m_{s'} \rangle$ is a measure of the temporal correlation between meta-orders and future collective latent order 
moves. Let us insist that we do not assume any causality here: $C(s-s')$ can be interpreted either 
as the information content of the order that {\it predicts} future price moves (i.e. the so-called ``alpha''), or as the collective reaction of the market to the order flow, i.e. the fact that agents may change their valuation as a result of the trading itself (see \cite{Bouchaud:2008,Bouchaud:2010} for a discussion of this duality). 

The second term in the right hand side of Eq. (\ref{2terms}) corresponds to the ``mechanical'' component of the impact discussed above, corresponding 
to the square-root impact. The first term, on the other hand, may behave very differently as a function of $T$. For example, if $C(s-s')$ has a range 
much smaller than $T$, the first term is expected to grow like $Q$ and not $\sqrt{Q}$.

\begin{figure}[h]
    \begin{center}
        \includegraphics{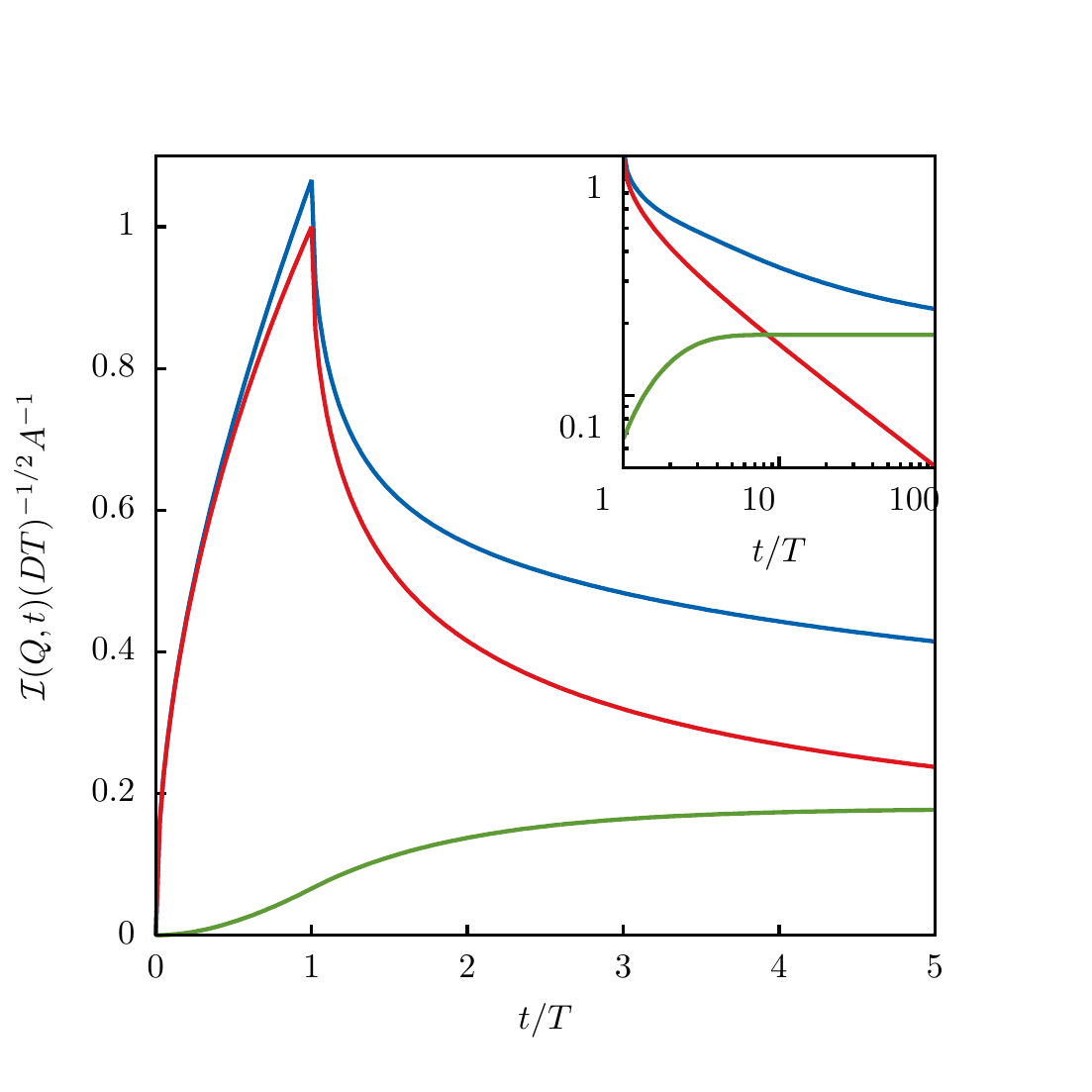}
    \end{center}
    \caption{The figure illustrates the relative r\^oles of mechanical and informational impact in determining the price 
    trajectory during and after the execution of a meta-order. 
    We have chosen in particular the set of parameters $D=J=\zeta=m=T=1$ and $\Gamma=0.1$. 
    The figure indicates that the mechanical part of the impact is the dominating effect at small times. 
    The permanent, informational component of the impact becomes relevant only after the slow decay of the mechanical component, 
    as shown the inset. From a theoretical point of view, the permanent component is important since it counterbalances potential
    market-making/mean-reversion profits coming from the confining effect of the latent order book on the price.}
    \label{fig:perm_vs_trans}
\end{figure}

When $t > T$, i.e. after the end of the meta-order, the informational contribution adds to the impact decay computed above and can substantially change the 
apparent evolution of $\langle p_t - p_0 \rangle$. In order to fix ideas, let us assume that $C(s-s')=\Gamma \zeta e^{-\zeta(s-s')}$ (other functional forms would not change the qualitative conclusions below). The behaviour of 
the  ``total'' impact for $t > T$ is then given by:
\be
{\cal I}_{\text{tot.}}(Q, t > T) = {\cal I}(Q, t > T) +  \Gamma Q - \frac{m_0 \Gamma}{ \zeta} 
(1 - e^{-\zeta T}) e^{-\zeta(t-T)} \underset{t  \to \infty}{\longrightarrow}  \Gamma Q,
\ee
which shows that on top of the relaxing mechanical impact (the first term), there is a growing contribution coming from the informational content of the 
trade (or alternatively from the collective reaction of the market to that trade) that saturates at large time to a finite value proportional to $Q$ -- see Fig. \ref{fig:perm_vs_trans}. 
This corresponds to a ``permanent'' component of impact.  
That the permanent component of impact should be linear in $Q$ conforms well with the assumptions of \cite{Kyle:1985,Almgren:2005}. 
However, our calculation shows that the empirical determination of the mechanical component of impact should carefully take into account 
any possible information content of the analyzed trades, as well as the possible auto-correlation of the trades. 
This parallels the discussion offered in \cite{Gomes:2013,Brokmann:2014}, where attempts are made to measure the decay of 
mechanical impact ${\cal I}(Q, t > T)$ in equity markets, with the conclusion that the mechanical component of impact seems indeed to relax to zero at large times. 

The possibility of generating a permanent impact by correlating the collective drift $V_t$ with the flow of meta-orders $m_s$ is 
in fact important to make our model internally consistent. Absent the permanent impact component, a random flow of meta-orders would 
give rise to a strongly mean-reverting contribution to the price (on top of the random walk contribution $\widehat p_t = \int_0^t {\rm d}s \, V_{s}$), 
and therefore potentially profitable mean-reversion/market making strategies. This profit can however be reduced to zero by increasing the 
permanent impact component (i.e. the $\Gamma$ factor above), that acts as an adverse selection bias for market makers. On this point, see the discussion in \cite{Wyart:2008}.

\section{Possible extensions and open problems}

The LLOB framework presented above is surprisingly rich and accounts for many empirical observations, but it can only be a first approximation of a more complex reality. First, we have
neglected effects that are in principle contained in Eqs.~(\ref{eq:dynamics}-a,b) but that disappear in the limit of slow latent order books $\nu T \ll 1$ 
(such that the memory time $1/\nu$ is much longer than the meta-order) and large liquidity ${\cal L}$, such that the meta-order only probes the linear
region of the book. Re-integrating these effects perturbatively is not difficult; for example, one finds that the impact ${\cal I}(Q)$ of a meta-order of size $Q$, 
executed at a constant rate, is lowered by a quantity proportional to $\nu T$ when $\nu T \ll 1$. In the opposite limit $\nu T \gg 1$, one 
expects that ${\cal I}(Q)$ becomes linear in $Q$, since impact must become additive in that limit (see~\cite{Toth:2011}). Any other large scale regularisation of the
model will lead to the same conclusion. One also expects that the deposition, with rate $\lambda$, of new orders behind the moving price should reduce the asymmetry of 
impact when the trade is reversed. We leave a more detailed calculation of these effects for later investigations.

Another important line of research is to understand the corrections to the LLOB induced by fluctuations, that are of two types: first, as discussed
in section \ref{dynamics}, the theory presented here only deals with the {\it average} order book $\varphi(x,t)$, from which the price $p_t$ is deduced using the definition
$\varphi(p_t,t)=0$, that allowed us to compute the average impact of a meta-order. However, one should rather compute the impact from the 
instantaneous definition of the price $p_t^{\text{inst.}}$ (that takes into account the fluctuations of the order book) and then take an average that would lead to ${\cal I}(Q)$. 
The numerical simulations shown in \cite{Iacopo:2014} suggest however that the approximation used here is quite accurate for long meta-orders, which is indeed 
expected as the difference 
$|p_t^{\text{inst.}}- p_t|$ becomes small compared to ${\cal I}(Q)$. 

Second, we have assumed that the rest of the market is in its stationary state and does not contribute to the source term modelling the meta-order. One should rather 
posit that the flow of meta-order $m_s$ has a random component that adds to the particular meta-order that
one is particularly interested in. There again, a calculation based on the average order book is not sufficient, since the interaction with other
uncorrelated meta-orders then trivially disappears. Following \cite{Cardy:1996}, one finds that random fluctuations in $m_s$ do contribute to a strongly mean-reverting term in the
variogram of the price trajectory, that should be taken into account in a consistent way.  Interestingly, this generic mean-reverting component leads to an excess short-term
volatility that is commonly observed in financial markets; more quantitative work on that front would therefore be worthwhile.

Finally, other extensions/modifications may be important in practice: as noted above, the cancellation rate $\nu$ is expected to increase with the
intensity of meta-orders. Furthermore, the incoming flow of latent orders $\lambda$ and/or the lifetime 
of orders $1/\nu$ can be expected to be increasing functions of the distance to the price $|x-p_t|$, i.e. better prices should attract more, and more patient buyers (or sellers), in such a way that the latent 
order book becomes {\it convex} at large distances. This would naturally explain why all impact data known to us
appear to grow even slower than $\sqrt{Q}$ at large $Q$ (\cite{Farmer:new, Lehalle:new}, and CFM, unpublished data). Another interesting path would be to allow the ``drift'' term 
$V_t$ in Eq.~(\ref{eq:dynamics}-a,b) to become non-Gaussian and thereby study a cumulant expansion of the square-root law.

\section{Conclusion}

In this paper, we have proposed a minimal theory of non-linear price impact based on a linear (latent) order book approximation, inspired
by diffusion-reaction models and general arguments. As emphasized in \cite{Toth:2011, Iacopo:2013, Iacopo:2014}, our modelling strategy does not rely 
on any equilibrium or fair-pricing conditions, but rather relies on purely statistical considerations. Our approach is strongly bolstered   
by the universality of the square-root impact law, in particular on the Bitcoin market -- as recently documented in \cite{Donier:2015} -- 
where fair-pricing arguments are clearly unwarranted because impact is much smaller than trading fees. 

Our framework allows us to compute the average price trajectory in the presence of a meta-order,
that consistently generalizes previously proposed propagator models. Our central result is the dynamical Eq.~(\ref{eq:price}), 
which not only reproduces the universally observed square-root impact law, 
but also predicts non-trivial trajectories when trading is interrupted or reversed. Quite surprisingly, 
we find that the short time behaviour of the free decay of impact is identical to that predicted by a propagator model, whereas the impact of a reversed trade is 
found to be much stronger. The latter result is in qualitative agreement with empirical observations \cite{Benichou}. We have shown that our model is free of price manipulation, which makes it
the first consistent, non-linear and time dependent theory of impact. Our setting also suggests how prices can be naturally decomposed into a transient ``mechanical impact'' component 
and a permanent ``informational'' component, as initially proposed by Almgren et al.~\cite{Almgren:2005}, and recently exploited in \cite{Gomes:2013,Brokmann:2014} -- see Section \ref{decomp}. 
Let us insist once again that this 
decomposition allowed us to construct {\it diffusive} prices (albeit with a generic short-term mean-reverting contribution).\footnote{A clear theoretical 
justification of the square-root impact law is also important if one wants to promote the idea of impact discounted mark-to-market accounting rules, 
as advocated in \cite{Caccioli}.}

Although our calculations are based on several approximations (restricting to a locally linear order book and
neglecting fluctuations), we believe that it provides a sound starting point for further extensions where the neglected
effects can be progressively reinstalled. Of particular importance is the potential feedback loop between price moves,
order flow and the shape of the latent and of the revealed order books. In particular, we have assumed that
latent orders instantaneously materialize in the real order book as the distance to the price gets small: any finite
conversion time might however contribute to liquidity droughts, in particular when prices accelerate, leading to an
unstable feedback loop. As emphasized in \cite{Lillo:2005, Bouchaud:2011, Toth:2011} this might be triggered by the anomalous liquidity fluctuations induced by the 
vanishingly small liquidity in the vicinity of the price.  This mechanism could explain the universal power-law distribution of returns that appear to be
unrelated to exogenous news but rather to unavoidable, self-induced liquidity crises.

\section*{Acknowledgments}

We warmly thank M. Abeille, R. Benichou, X. Brokmann, J. de Lataillade, C. Deremble, J. D. Farmer, J. Gatheral, J. Kockelkoren, C. A. Lehalle, 
Y. Lemp\'eri\`ere, F. Lillo, E. S\'eri\'e and in particular M. Potters and B. T\'oth for many discussions 
and collaborations on these issues. We also thank P. Blanc, N. Kornman, T. Jaisson, M. Rosenbaum and A. Tilloy for useful remarks on the manuscript.  
One of us (IM) benefited from the support of the ``Chair Markets in Transition'', under the aegis of ``Louis Bachelier Finance and Sustainable Growth'' laboratory, 
a joint initiative of \'Ecole Polytechnique, Universit\'e d'\'Evry Val d'Essonne and F\'ed\'eration Bancaire Fran\c{c}aise.

\section*{Appendix A: Derivation of the drift/diffusion term}

In order to give more flesh to the microscopic assumptions underlying the drift/diffusion equation written in Eqs.~(\ref{eq:dynamics}-a,b), 
let us assume first that each agent 
contributes to a negligible fraction of the latent order book, 
which is probably a good approximation for deep liquid markets. A model for thin markets, where some participants contribute to a substantial fraction of the liquidity, is discussed below, but 
leads to a very similar final result.

Between $t$ and $t + \delta t$, each agent $i$ revises its reservation price 
$p_i$ to $p_i + \beta_i \xi_t + \eta_{i,t}$, where $\xi_t$ is common to all $i$ representing some public information (news, but also the price change itself or the order flow, etc.) 
and $\beta_i > 0$ is the sensitivity of agent $i$ to the news, which we imagine to be a random variable from agent to agent, 
with a pdf $\Pi(\beta)$ mean normalized to $[\beta_i]_i=1$. [$[...]_i$ represents a cross-sectional average over agents.] Some agents may over-react, 
others under-react; $\beta_i$ might in fact be itself time dependent, but we assume that the {\it distribution} of $\beta$'s is stationary. 
The  completely idiosyncratic contribution $\eta_{i,t}$ is an independent random variable both across different agents and in time, 
with distribution $R(\eta)$ of mean zero and rms $\Sigma$. We assume that within each price interval $x, x+ {\rm d} x$ lie latent orders from a large number of agents. The density of latent orders $\rho(x,t)$ therefore evolves according to the following Master equation:
\be
\rho(x,t+\delta t) = \int_0^{\infty} {\rm d}\beta \Pi(\beta) \int_{-\infty}^{\infty} {\rm d}\eta R(\eta) \int {\rm d}y \rho(y,t)  \delta(x - y - 
\beta \xi_t - \eta),
\ee
or:
\be
\rho(x,t+\delta t) = \int_0^{\infty} {\rm d}\beta \Pi(\beta) \int_{-\infty}^{\infty} {\rm d}\eta R(\eta)  
\rho(x - \beta \xi_t - \eta,t).
\ee
Assuming that the price revisions $\beta \xi_t + \eta$ over a small time interval $\delta t$ are small enough, a second-order expansion Kramers-Moyal of the above equation leads to 
(see \cite{Gardiner} for an in-depth discussion of this procedure):
\be
\rho(x,t+\delta t) - \rho(x,t) = - \xi_t \rho'(x,t) + \frac12 \left([\beta^2]\xi_t^2 + \Sigma^2 \right) \rho''(x,t) + \dots
\ee
At this stage, one can either assume that formally $\xi_t = V_t \delta t$ and $\Sigma^2 = 2D_0 \delta t$ in which case the continuous time limit reads:
\be
\frac{\partial \rho(x,t)}{\partial t} = - V_t \frac{\partial \rho(x,t)}{\partial x} + D_0 \frac{\partial^2 \rho(x,t)}{\partial x^2}
\ee
or that $\xi_t = V_t \sqrt{\delta t}$, where $V_t$ is now a Gaussian white noise of variance $\sigma^2$, and again $\Sigma^2 = 2D \delta t$, in which 
case the continuous time limit should be written as:
\be\label{eq:Jonathan}
{\rm d}\rho(x,t) = - {\rm d}W_t \frac{\partial \rho(x,t)}{\partial x} +  D_1 {\rm d}t \frac{\partial^2 \rho(x,t)}{\partial x^2}
\ee
with ${\rm d}W_t$ a Wiener noise and $D_1 \equiv D_0 + [\beta^2] \sigma^2/2$, to wit, the diffusion constant involves both the idiosyncratic component and the dispersion of reaction to random information. 
This is the interpretation we will mostly follow in the present paper. A careful derivation of the
corresponding equation in the reference frame of the price $\widehat p_t = \int_0^t {\rm d}W_s$ finally gives the diffusion part of 
Eq.~(\ref{eq:diff_psi2}) in the main text:
\be\label{eq:J2}
\frac{\partial \rho(y,t)}{\partial t} =  D \frac{\partial^2 \rho(y,t)}{\partial y^2}; \qquad D \equiv D_0 + \frac{\sigma^2}{2} \int {\rm d}\beta \Pi(\beta) (\beta-1)^2;
\ee
i.e. only the dispersion of reaction $\beta-1$ can contribute to the diffusion term, as expected.

Another interpretation of this last equation is to imagine that between $t$ and $t+ {\rm d}t$, a fraction $\phi \in [0,1]$ (possibly time
dependent) of agents collectively change their price estimate by an amount ${\rm d}W_t$, with no other idiosyncratic component. This leads to:
\be
{\rm d}\rho(x,t) = \phi \left[\rho(x-{\rm d}W_t,t) - \rho(x,t) \right] = - \phi {\rm d}W_t \rho'(x,t) + \frac12 \phi \sigma^2 {\rm d}t \rho''(x,t),
\ee
that essentially corresponds to the case above with $\Pi(\beta)=(1 - \phi) \delta(\beta) + \phi \delta(\beta-1)$. In the price reference frame
$\widehat p_t = \int_0^t \phi {\rm d}W_s$, one finds Eq.~(\ref{eq:J2}) with $D \equiv \phi(1-\phi) \sigma^2/2$. Note that, clearly, these collective
price revisions must by themselves induce transactions whenever $0 < \phi < 1$.

If price revisions cannot be considered as small, the resulting evolution of $\rho(x,t)$ should include jumps in the continuous time limit, i.e. 
one would find an integro-differential equation rather than a partial differential equation for $\rho(x,t)$. However, if the jump process is 
homogeneous in space, one can diagonalize the evolution operator in Fourier space. This allows one to show that price manipulation is impossible in that
case as well.

\section*{Appendix B: A generically linear latent order book}

Let us consider the case where the deposition flow is not constant. This leads to the following equation for the stationary state of the latent 
order book:
\begin{equation}
\label{eq:diff_psi4}
D \frac{\partial^2 \varphi_{\text{st.}}(y)}{\partial y^2} - \nu \varphi_{\text{st.}}(y) + \lambda \, (\Theta(y)-\Theta(-y)) =0,
\end{equation}
with $\varphi_{\text{st.}}(y) = - \varphi_{\text{st.}}(-y)$ (and in particular the market clearing condition $\varphi_{\text{st.}}(y=0)=0$).

Let us assume that $\Theta(y)-\Theta(-y)$ behaves, for $y \to \infty$, as a constant that we can set to unity. The solution $\varphi_{\text{st.}}(y)$ then converges to $\lambda/\nu$ for large $y$, so we set:
\be
\varphi_{\text{st.}}(y) = \frac{\lambda}{\nu} + \Psi(y),
\ee
where $\Psi(y)=1 - \Theta(y) + \Theta(-y)$ with $\Psi(y \to \infty) \to 0$, and:
\begin{equation}
\label{eq:diff_psi5}
D \frac{\partial^2 \Psi(y)}{\partial y^2} - \nu \Psi(y) = \lambda \Xi(y),
\end{equation}
where $\Xi(y \to \infty) \to 0$. The boundary condition on $\Psi(y)$ at large $y$ means that we can look at a solution of the form:
\be
\Psi(y) = \psi(y) e^{-\sqrt{\nu/D}\, y},
\ee
so that:
\be
D \psi''(y) - 2 \sqrt{\nu D} \psi'(y) = \lambda \Xi(y) e^{\sqrt{\nu/D}\, y}.
\ee
Finally, one finds:
\be
\varphi_{\text{st.}}(y) = \frac{\lambda}{\nu} \left[1 - e^{-\sqrt{\nu/D}\, y}\right] + \frac{\lambda}{D}  e^{-\sqrt{\nu/D}\, y} \int_0^y {\rm d}y'
 e^{2\sqrt{\nu/D} \,y'} \int_{y'}^\infty {\rm d}y''
 e^{-\sqrt{\nu/D} \,y''}\Xi(y'').
\ee
The solution given in the main text corresponds to $\Xi \equiv 0$, so that only the first term survives. From the above explicit form, one sees that provided the integral 
$\int_{0}^\infty {\rm d}y'' e^{-\sqrt{\nu/D} y''}\Xi(y'')$ is finite, the behaviour of $\varphi_{\text{st.}}(y)$ is {\it always} linear in the 
vicinity of $y =0$. Only a highly singular the deposition rate, diverging faster than $1/y$ when $y \to 0$, would jeopardize
the local linearity of the latent order book (see also \cite{Bouchaud:2002}, where this property was first discussed).

\section*{Appendix C: Shape of the order book during constant rate execution and initial relaxation of impact}

When the trading rate is a constant (equal to $m_0$),  one can exhibit an {\it exact} scaling solution of the time dependent order book of the form 
$\varphi(x,t)=m_0 \sqrt{t/D}\, F(\frac{x}{\sqrt{Dt}})$, where $F$ is the solution of:
\be
2F''(u) + u F'(u) - F(u) = - 2\delta(u - A).
\ee
As we show below, this equation can be solved and gives the exact shape of the book at $t=T$, from which the initial relaxation (after trading has stopped) can be deduced. 

Writing $F=u G$ in the above equation, one finds a first order linear equation for $H=G'$:
\be
H'(u) + \left(\frac{u}{2} + \frac{2}{u}\right) H(u) =  \frac{1}{u} \delta(u - A)
\ee
which is easily solved as:
\be
H(u) = \frac{H_0}{u^2} e^{-u^2/4}, \quad (u < A);\qquad  H(u) = \frac{H_0 - A e^{A^2/4}}{u^2} e^{-u^2/4}, \quad (u > A).
\ee
There are two boundary conditions that are useful. One is the very definition of the price position, $x = A \sqrt{Dt}$ or $A=u$, for
which $\phi(x,t) = x$ or $F(A)=AJ/m_0$. The second remark is that when $u=0$, the integral defining $F$ can be computed, leading to
\be
F(0) = \frac{A}{4 \sqrt{\pi}} e^{A^2/4} \int_{A^2/4}^\infty \frac{{\rm d}v}{v^{3/2}} e^{-v}.
\ee
This allows one to fix $H_0$ since $G'(u)=F'(0)/u - F(0)/u^2 \approx -F(0)/u^2$ when $u \to 0$, to be compared with $H(u) \approx H_0/u^2$ in the
same limit. Hence $H_0 = - F(0)$. 

\begin{figure}[h]
    \begin{center}
        \includegraphics{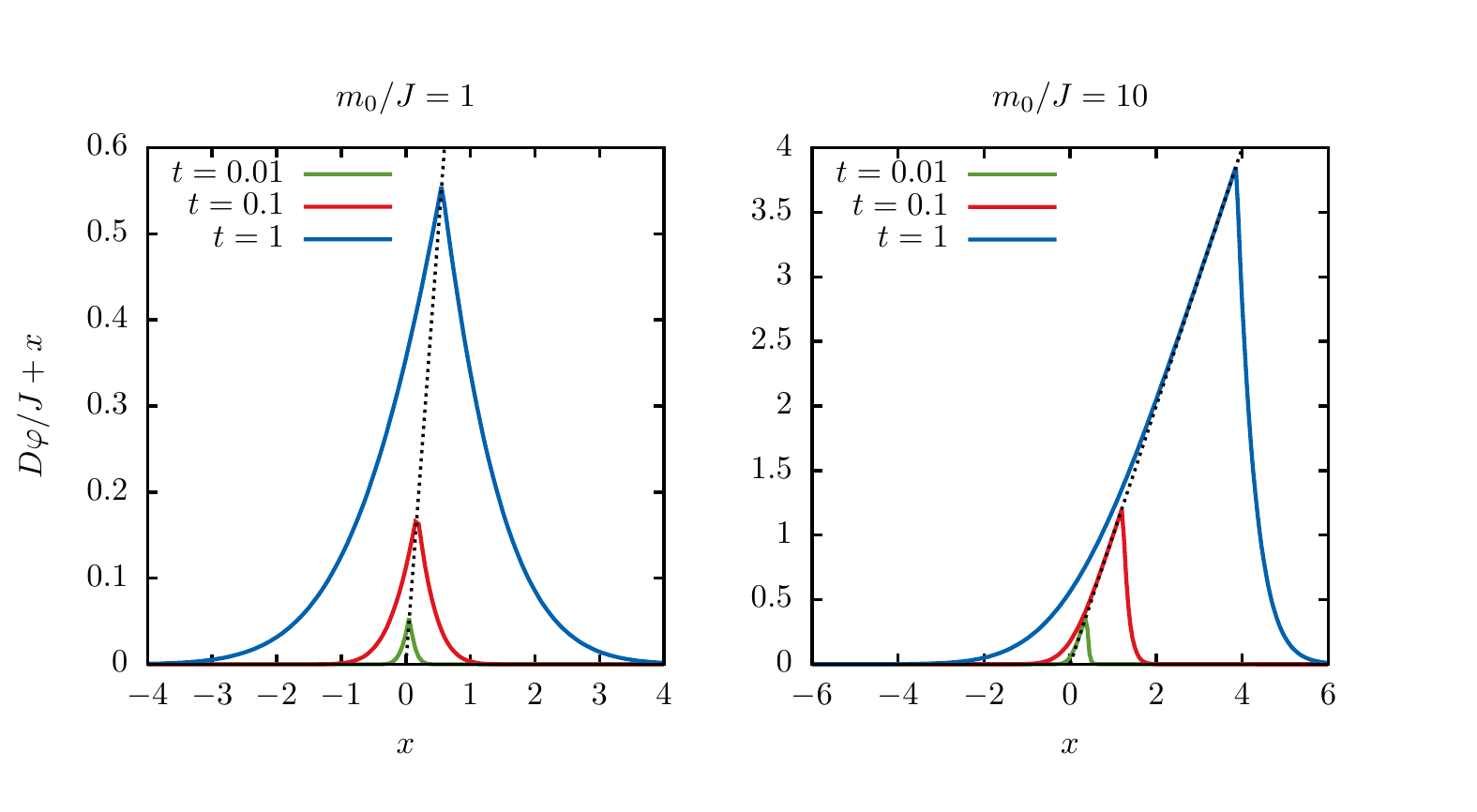}
    \end{center}
    \caption{Evolution of the average order book $\varphi(x,t)$, represented for two different values of the perturbation parameter $m_0/J$. 
    The curves in different colors are snapshots taken at different times of the difference between the perturbed and the unperturbed average of the book. 
    The scaling form of the book $\varphi(x,t)=m_0 \sqrt{t/D} F(\frac{x}{\sqrt{Dt}})$ is clear from the plots, which illustrate how a flat region of the book is formed at large values of $m_0/J$. }
    \label{fig:book_evol}
\end{figure}

The final solution for $F(u)$ is easily obtained from integrating $H$ and multiplying by $u$ (see Fig.~\ref{fig:book_evol}). Using $G(A)=J/m_0$, 
one finds:
\be
G(u) = F(0) \int_u^A \frac{{\rm d} v}{v^2} e^{-v^2/4} + J/m_0, \qquad (u \leq A)
\ee
and
\be
G(u) = - (F(0) + A e^{A^2/4}) \int_A^u \frac{{\rm d} v}{v^2} e^{-v^2/4} + J/m_0, \qquad (u \geq A)
\ee
Of special interest is the slope of $F$ for $u=A^{\pm}$; with $F'=uH(u)+G(u)$ one finds:
\be
F'(A^-) = J/m_0 - \frac{F(0)}{A} e^{-A^2/4}; \qquad F'(A^+)= J/m_0 - \frac{F(0)}{A} e^{-A^2/4} - 1.
\ee

Now, shortly after the meta-order has stopped, one can look at the solution of the diffusion equation in the vicinity of the 
final price $p_T=A  \sqrt{DT}$, using a piece-wise linear function for the initial condition, with slopes given by $F'(A^\pm)$.  
The solution then reads, with $t - T = \Delta$ small:
\be
\varphi(x,t) = m_0 \sqrt{T \Delta} \left[ F'(A^-) z + (F'(A^+)-F'(A^-)) \int_z^{\infty} {\rm d} u \frac{(u-z)}{\sqrt{4\pi}} e^{-u^2/4} 
\right],
\ee
with $z = (p_T - x)/\sqrt{D \Delta}$. The position of the price is given by $p_t = p_T - \sqrt{D \Delta} z^*$, with $z^*$ such that:
\be
F'(A^-) z^* + (F'(A^+)-F'(A^-)) \int_{z^*}^{\infty} {\rm d} u \, \frac{(u-z^*)}{\sqrt{4\pi}} e^{-u^2/4}=0.
\ee
Using the expression for $F(0)$ and the result above for $F'(A^\pm)$, this equation simplifies to:
\be
z^* \int_{A^2/4}^\infty \frac{{\rm d}v}{v^{3/2}} e^{-v} = 2 \int_{z^*}^{\infty} {\rm d} u \,(u-z^*) e^{-u^2/4}.
\ee
Changing variables in the RHS from $u$ to $v=u^2/4$, and integrating by parts, one finds:
\be
z^* \int_{A^2/4}^\infty \frac{{\rm d}v}{v^{3/2}} e^{-v} = z^* \int_{z^{*2}/4}^{\infty} \frac{{\rm d}v}{v^{3/2}} e^{-v},
\ee
which leads to $z^* = A$ for all $m_0/J$. [The other solution, $z^*=0$, is spurious]. 

Hence, the initial stage of the impact relaxation can be written in a super-universal way:
\be
p_t \underset{t \to T^+}{\approx} p_T \left[1 - \sqrt{\frac{t-T}{T}}\right],
\ee
i.e. exactly the result from the propagator model, even in the non-linear regime!

\section*{Appendix D: A saddle point approximation for large trading rates}
\label{sec:a_saddle_point_approximation_for_large_trading_rates}

In this Appendix we develop a systematic procedure in order to solve perturbatively Eq.~\eqref{eq:price}. The first step in order to find a solution is to introduce an expansion parameter 
$\epsilon \ll 1$, which we use in order to control the amplitude of the trading rate through $m_t \to m_t \epsilon^{-1}$. Such substitution implies a scaling of the solution of the form $y_t \to y_t \epsilon^{-1/2}$,
leading to an equation for the price of the form:
\begin{equation}
{\cal L} y_t  = \int_0^t \frac{{\rm d}s\, m_s}{\sqrt{4 \pi D\epsilon  (t-s)}} e^{-\frac{(y_t - y_s)^2}{4D\epsilon (t-s)}}.    \,
\end{equation}
which is equivalent to the one which one would have by leaving invariant $m_t$ and by performing the substitutions $D \to D\epsilon$ and $J \to J\epsilon$. 
Hence, the large trading regime is equivalent to the one of slow diffusion.

In this case, it is evident that the integral is dominated by times $s$ close to $t$, which suggests to Taylor expand both the trading rate $m_s$ and the price $y_s$ around $s=t$, 
so to insert the resulting series in the integral appearing in Eq.~\eqref{eq:price}.
The dominating term results
\begin{equation}
    \label{eq:exp_leading}
    m_t \int_0^\infty {\rm d}u \frac{1}{\sqrt{4\pi D u \epsilon}} e^{-\dot y^2_t \frac{ u}{4D\epsilon}} =
    m_t|\dot y_t|^{-1} \;.
\end{equation}
The successive corrections to above result can be computed systematically, as they involve developing the square and the exponential function in the Gaussian term in Eq.~\eqref{eq:price}. In particular, by exploiting the identity
\begin{equation}
    \int_0^\infty {\rm d}u\,e^{-z^2 u} u^\alpha = \Gamma(1+\alpha) |z|^{-2(1+\alpha)} \,
\end{equation}
it is possible to derive the expansion
\begin{eqnarray}
    \label{eq:final_exp}
    {\cal L} y_t |\dot y_t | &=& m_t \bigg[ 1
    + (D\epsilon) \left(\frac{3 {\ddot y_t}}{{\dot y_t}^3}-\frac{2 {\dot m_t}}{{m_t} {\dot y_t}^2}\right) \\
    \nonumber & + &  (D\epsilon)^2 \left( \frac{ 6 {\ddot m_t} {\dot y_t}^2-30 {\dot m_t} {\ddot y_t} {\dot y_t}-10 {m_t} {\dddot{ y_t}} {\dot y_t}+45 {m_t} {\ddot y_t}^2}{{m_t} {\dot y_t}^6}\right) \\
    \nonumber &+& 5(D\epsilon)^3 \left( \frac{ -4 {\dddot{m_t}} {\dot y_t}^3+42 {\ddot m_t} {\ddot y_t} {\dot y_t}^2+28 {\dot m_t} {\dddot{ y_t}} {\dot y_t}^2+7 {m_t} {\ddddot{y_t}} {\dot y_t}^2}{{m_t} {\dot y_t}^9} \right) \\
    \nonumber &+& 5(D\epsilon)^3 \left( \frac{-168 {\dot m_t} {\ddot y_t}^2 {\dot y_t}-112 {m_t} {\ddot y_t} {\dddot{ y_t}} {\dot y_t}+252 {m_t} {\ddot y_t}^3}{{m_t} {\dot y_t}^9} \right) \\
    \nonumber &+& O(\epsilon^4)
    \bigg] \, ,
\end{eqnarray}
whose first-order terms match the form reported in Eq.~\eqref{eq:expansion}. Each of the contributions of order $\epsilon^n$ can be seen equivalently either as suppressed by 
the small value of the diffusion constant diffusion (through a $D^n$ factor) or by the large value trading rate (through a factor of the order of $m_t^{-n}$).

Finally, note that the implicit equation above needs to be inverted in order to obtain a relation yielding $y_t$ as a function of $m_t$. 
This is possible by using Eq.~\eqref{eq:final_exp} as an iterative scheme for $y_t$, which allows to calculate
\begin{eqnarray}
    {\cal L} y_t |\dot y_t|& =& m_t \\
    \nonumber &+& (J\epsilon) \left( -3 + \frac{2 Q_t {\dot{m_t}}}{m^2_t}\right) \\
     \nonumber &+& (J\epsilon)^2\left( -\frac{12 Q_t {\dot{m_t}} }{m_t^3} - \frac{6 \, s\, {\dot{m_t}} }{m_t^2} + 
     \frac{4{\dot{m_t}}}{m_t^2}\int {\rm d}s \frac{Q_s{\dot{m_s}}}{m^2_s} + \frac{16 Q^2_t {\dot{m_t}}^2}{m_t^5} - \frac{4 Q^2_t {\ddot{m_t}}}{m_t^4}\right) \\
     \nonumber &+& O(\epsilon^3) \, ,
\end{eqnarray}
where $Q_t = \int_0^t {\rm d}s \, m_s$.

As a simple application of the above formula, consider the case where $m_t \geq 0, \forall t \in [0,T]$. In this case, $\dot y_t$ is also non negative and we can remove the absolute value in the
above equation. To order $\epsilon$, the solution of the above equation is (assuming $y_0 = 0$):
\be
\frac12 {\cal L} y_t^2 = Q_t + J \epsilon \,(- t - \frac{2 Q_t}{m_t}),
\ee
where we have used integration by parts. Now, the cost ${\cal C}(Q)$ associated to buying a total quantity $Q$ in time $T$ is given by:
\be
{\cal C}(Q) = \int_0^T {\rm d}s \, m_s y_s, \qquad Q =  \int_0^T {\rm d}s \, m_s.
\ee
Therefore, to order $\epsilon$:
\be
{\cal C}(Q) = \sqrt{\frac{2}{\cal L}} \int_0^T {\rm d}s \, m_s \sqrt{Q_s} - \frac{J \epsilon}{\sqrt{2 {\cal L}}} \int_0^T {\rm d}s \, \left[2\sqrt{Q_s} + \frac{s m_s}{\sqrt{Q_s}}\right].
\ee
After integrating by parts the last term, one finally finds that the impact cost is, to order $\epsilon$, {\it independent} of the trading schedule $m_s$, and given by:
\be
{\cal C}(Q) = \frac23 \sqrt{\frac{2}{\cal L}} Q^{3/2}\left[  1 - \frac{3 J \epsilon T}{2Q}  \right].
\ee
The correction term is negative, as expected since a slower trading speed leaves time for the opposing liquidity to diffuse towards the traded price.

\end{document}